\documentclass[traditabstract]{aa} 
\usepackage{amsmath}
\usepackage{graphicx}
\usepackage{txfonts}
\usepackage{comment}
\usepackage{natbib}
\bibpunct{(}{)}{;}{a}{}{,}
\usepackage{float}
\usepackage{calc}
\usepackage{textcomp}
\usepackage{placeins}
\usepackage{color}
\usepackage{rotating} 
\usepackage{lscape}
\usepackage{url}
\usepackage[colorlinks=true,linkcolor=blue,citecolor=blue]{hyperref}
\usepackage{subfig}
\usepackage[all]{draftcopy}
\usepackage{longtable}
\usepackage{array}
\usepackage{threeparttable}
\usepackage{lscape}
\usepackage{enumerate}
\DeclareTextSymbol{\degre}{OT1}{23}
\newcounter{savedfootnote}

\def \cigale{{{\sc cigale}}}
\renewcommand{\epsilon}{\varepsilon} 

\begin{document}
\title{Identification of a transition from stochastic to secular star formation around $z=9$ with JWST.}

\author{
L.~Ciesla\inst{1}\fnmsep\thanks{\email{laure.ciesla@lam.fr}},
D.~Elbaz\inst{2},
O.~Ilbert\inst{1},
V.~Buat\inst{3},
B.~Magnelli\inst{2},
D.~Narayanan\inst{4,5},
E.~Daddi\inst{2},
C.~G{\'o}mez-Guijarro\inst{2},
and R.~Arango-Toro\inst{1}.
}

\institute{	
Aix Marseille Univ, CNRS, CNES, LAM, Marseille, France
\and
Université Paris-Saclay, Université Paris Cité, CEA, CNRS, AIM, 91191, Gif-sur-Yvette, France
\and
Aix-Marseille Univ., CNRS, CNES, LAM, 13388, Marseille Cedex 13, France; Institut Universitaire de France (IUF), Paris, France
\and
Department of Astronomy, University of Florida, 211 Bryant Space Sciences Center, Gainesville, FL 32611, USA
\and
Cosmic Dawn Center at the Niels Bohr Institute, University of Copenhagen and DTU-Space, Technical University of Denmark, Denmark
}		

   \date{Received; accepted}

  \abstract
{
Star formation histories (SFH) of early (6$<z<$12) galaxies have been found to be highly stochastic in both simulations and observations, while at $z\lesssim$6 the presence of a main sequence (MS) of star-forming galaxies imply secular processes at play. 
In this work, we aim at characterising the SFH variability of early galaxies as a function of their stellar mass and redshift.
We use the JADES public catalogue and derive the physical properties of the galaxies as well as their SFH using the spectral energy distribution modelling code {\sc cigale}.
To this aim, we implement a non-parametric SFH with a flat prior allowing for as much stochasticity as possible.
We use the SFR gradient, an indicator of the movement of galaxies on the SFR-$M_\ast$ plane, linked to the recent SFH of galaxies.
This dynamical approach of the relation between the SFR and stellar mass allows us to show that, at $z>9$,  87\% of massive galaxies, ($\log(M_\ast/M_\odot)\gtrsim$9), have SFR gradients consistent with a stochastic star-formation activity during the last 100\,Myr, while this fraction drops to 15\% at $z<7$.
On the other hand, we see an increasing fraction of galaxies with a star-formation activity following a common stream on the SFR-$M_\ast$ plane with cosmic time, indicating that a secular mode of star-formation is emerging.
We place our results in the context of the observed excess of UV emission as probed by the UV luminosity function at $z\gtrsim10$, by estimating $\sigma_{UV}$, the dispersion of the UV absolute magnitude distribution, to be of the order of 1.2\,mag and compare it with predictions from the literature. 
In conclusion, we find a transition of star-formation mode happening around $z\sim9$: 
Galaxies with stochastic SFHs dominates at $z\gtrsim9$, although this level of stochasticity is too low to reach those invoked by recent models to reproduce the observed UV luminosity function.
}

   \authorrunning{Ciesla et al.}
   \titlerunning{Transition between stochastic and secular star-formation modes at z$\sim$9}

   \maketitle

\section{\label{intro}Introduction}

Since the Universe was 1\,Gyr, the majority of galaxies have formed their stars via secular process, as shown by the existence of the main sequence (MS) of star-forming galaxies \citep{Noeske07_SFseq,Elbaz07,Daddi07}.
Before JWST, at $z\gtrsim$7 only the rest-frame optical range was accessible thanks to \textit{Spitzer}, in which strong nebular lines can mimic the presence of a Balmer break and bias the analysis \citep[e.g.,][]{Labbe13,Smit14,RobertsBorsani16,DeBarros19,Tang19}.
Simulations predict that star formation in early galaxies increases with time during the epoch of reionization (EoR), transitioning from being stochastic to continuous as their gravitational potentials become deep enough to withstand supernovae and radiative feedbacks \citep{Dayal13, KimmCen14, Faisst19, Emami19, Wilkins23}. 
This transitional stellar mass, between the stochastic SFH and the continuous phase, is estimated around $\log$($M_\ast$/M$_{\odot})$$=$8.
Indeed, the fraction of stellar mass formed while in stochastic phase decreases with redshift and increasing stellar mass: galaxies with a stellar mass of $\log$($M_\ast$/M$_{\odot})$$\sim$7 have spent about 70$\%$ of their lifespan in stochastic phase at $z$$=$5, defined as being located more than 0.6\,dex away from the average SFR-$M_\ast$ relation  \citep{Ma18,Legrand22}.
Using a sample of local dwarf galaxies, analogues of high redshift objects, \cite{Emami19} showed that the distribution of the H$\alpha$ to ultraviolet (UV) luminosity ratio, a proxy for burstiness of the SFH, is wider in low mass galaxies ($<$10$^8$\,M$_{\odot}$) than in higher mass galaxies, indicating this transition mass.

Furthermore, a wide range of star formation activity is discovered in these early galaxies showing how stochastic, or bursty, the SFH is at these redshifts \citep{Endsley23,Dressler23b,Looser23b}.
Early galaxies are found in different phases of their star-formation (SF) activity when observed in spectroscopy with NIRSpec.
\cite{Looser23b} showed examples of galaxies that they classify as experiencing a regular phase, where emission lines are strong and clearly observed. 
Galaxies are found in a ``lull" phase where emission lines are present but weak. 
And finally, others can be found with no emission lines, in a ``mini-quenched" phase, as proposed by the authors, or in a state of inactivity. 
As an example, \cite{Looser23a} discovered an inactive galaxy at $z=7.3$ with a stellar mass of $\log$($M_\ast$/M$_{\odot})$$=$8.7.
SED modelling showed that this galaxy experienced a short and intense burst of star formation followed by rapid quenching, about 10 to 20\,Myr before the epoch of observation.
Since supernova feedback can only act on a longer timescale ($>$30\,Myr), \cite{Gelli23} argued that the observed abrupt quenching must be caused by a faster physical mechanism such as radiation-driven winds.
Galaxies in this low mass range ($\log$($M_\ast$/M$_{\odot})$$\sim$8) are indeed sensitive to various feedback mechanisms that can result in temporary or permanent quiescence.
\cite{Gelli23} built a sample of 130 low mass ($\log$($M_\ast$/M$_{\odot})$$<$9.5) galaxies from the {\sc serra} cosmological zoom-in simulation and found that feedback regulates their bursty SFHs.
They estimated that, on average, 30\% of the galaxies in their sample are quiescent in the $z$$=$6 to 8.4 redshift range and are the dominant population at $M_\ast$ lower than $\log$($M_\ast$/M$_{\odot})$$=$8.3.

This stochasticity of the SFH of galaxies is mentionned as one of the possibility that could explain the UV excess of the UV luminosity function (UVLF) at high redshifts ($z$$\gtrsim$10).
Indeed, with the launch of JWST, several hundreds of galaxies have been observed at redshifts higher than 9 \citep[e.g.,][]{Donnan23,Naidu22,Finkelstein23,Qin23,Harikane23,Rieke23} and an overabundance of UV bright galaxies is observed resulting in little evolution of the UVLF at $z$$\gtrsim$10 \citep[e.g.,][]{Harikane23,Finkelstein23}.
Confirmed with spectroscopic samples \citep[e.g.,][]{Arrabal23,CurtisLake23,Robertson23,Harikane23b}, several physical explanations are brought forward to understand the UVLF tension between observations and models \citep[e.g.,][]{Inayoshi22,Ferrara22,Dekel23,Mason23,Yung23}.
These explanations aimed at increasing the median UV radiation emitted by early galaxies (higher star-formation efficiency, a top-heavy stellar initial mass function, absence of dust attenuation, non-stellar UV emission sources such as black holes).
Another approach is to increase the stochasticity of star-formation, that is galaxies experiencing a high star formation activity will be brighter in UV. 
Therefore, because of the steep decline of the halo mass function in the bright end of the UVLF, there are more intrinsically low-mass sources upscattered to high luminosities than massive galaxies downscattered to faint luminosities, which will populate the bright end of the UVLF \citep[e.g.,][]{Shen23}.

In this paper, we aim at understanding how the hints of (i) the presence of a MS at 6$<z<$12 and (ii) highly stochastic SFHs found in $\log$($M_\ast$/M$_{\odot})$$<$9 galaxies can be reconciled, as well as characterising the stochasticity of galaxies' SFH.
We use JADES \citep{Rieke23} early galaxies ($z>6$) and place them on the SFR-$M_\ast$ mass plane to discuss the existence of a MS in the first Gyr of the Universe.
To do so, we implement a non-parametric SFH with a flat prior, tested and calibrated on simulations.
Using the SFR gradient, a SED modelling output indicating the movement of galaxies in the SFR-stellar mass plane introduced and tested by \cite{Ciesla23} and \cite{ArangoToro23}, we investigate the burstiness of early galaxies' SFH.
We present the sample in Sect.~\ref{sample}.
The new non-parametric SFH for {\sc cigale} is presented and tested in Sect.~\ref{sedfit}.
The presence of a relation between the SFR and stellar mass at $z>6$ is discussed in Sect.~\ref{secms} while the burstiness of the SFH is investigated in Sect.~\ref{sectrans}.
Discussion and conclusions are presented in Sect.~\ref{discussion} and Sect.~\ref{conclusions}, respectively.
Throughout the paper, we use a \cite{Salpeter55} initial mass function and WMAP7 cosmology \citep{Komatsu11}.
\section{\label{sample}The sample}

In this work, we use the JWST Advanced Deep Extragalactic Survey \citep[JADES,][]{Bunker23,Eisenstein23,Hainline23} initial data release of the Hubble Ultra Deep Field covering 26$\arcmin^2$ presented in \cite{Rieke23}.
This catalogue combines 9 broad filters from JWST/NIRCam, 5 NIRCam medium bands from the JEMS survey \citep{Williams23}, and existing HST imaging, for a total of 23 photometric bands.
The reddest NIRCam filter where all the galaxies are detected (F444W) allows us to probe the 0.63\,$\mu$m and 0.34\,$\mu$m rest frame at $z=6$ and $z=12$, respectively.
The catalogue includes sources detected in the F200W NIRCam filter and photometry performed in a set of apertures.
In the wide bands, 5$\sigma$ flux depths of the 9 NIRCam bands are between 3.4 and 5.9\,nJy, while they range from 6.1 to 18.8\,nJy in the medium bands.
Photometric redshifts available in the catalogue are computed with the code \texttt{EAZY} \citep{Brammer08}.
The average offset between photometric redshifts and a compilation of spectroscopic redshifts is 0.05 with a $\sigma$ of 0.024.
We use spectroscopic redshifts when available although it concerns only a few galaxies.
We select galaxies with $z>6$ in the JADES catalogue and use the photometry performed using KRON apertures.
We refer the reader to \cite{Rieke23} for more details on the data reduction process, photometry procedures, and redshift estimates.

\begin{figure}[ht] 
 	\includegraphics[width=\columnwidth]{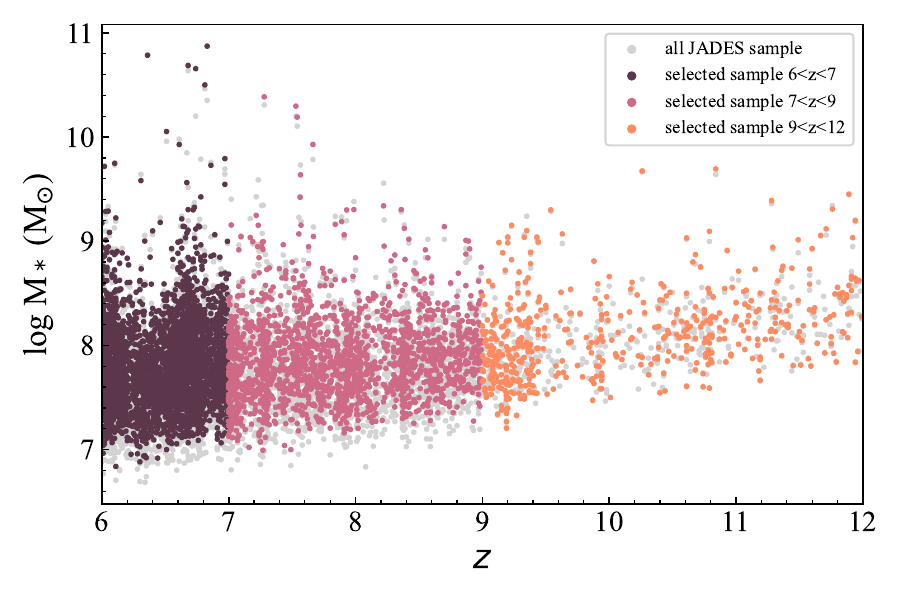}
  	\caption{\label{zdist} Stellar mass as a function of redshift for the JADES sample (grey dots) and the selected galaxies studied in this work. Colours highlight the three redshift bins analysed in this work.}
\end{figure}

Our sample comprises 5,601 galaxies, for which we show the stellar mass of galaxies as a function of their redshift in Fig.~\ref{zdist}.
Selected galaxies have at least 2 JWST bands with an SNR larger than 3.
We separate the galaxies in three redshift bins: $6<z<7$, $7<z<9$, and $9<z<12$, spanning 150, 220, and 180\,Myr, respectively.
They include 3,197, 1,883, and 521 galaxies, respectively.

\section{\label{sedfit}A non-parametric SFH tailored for early galaxies.}

Physical properties and SFH reconstruction are obtained through SED modelling of the JADES 23 photometric bands using the code {\sc cigale}\footnote{\url{https://www.lam.fr/}} \citep{Boquien19} to which we include a non-parametric SFH module adapted for galaxies at high redshifts.

\subsection{\label{cigale}The SED modelling procedure}
{\sc cigale} is a public and versatile code that models galaxies' SED from UV to radio taking into account the balance between the energy absorbed by dust in the UV-optical and reemitted in the IR.
The code can be used to fit photometry and spectroscopy, or to create mock SEDs thanks to its large library of models.
{\sc cigale} has been extensively tested and a comparison with other SED modelling codes can be found in \cite{Pacifici23}.
In this work, we use a non-parametric SFH assumption, the stellar population models of \cite{BruzualCharlot03}, and a \cite{Calzetti00} dust attenuation law which is found to be suitable for early galaxies \citep{Bowler22}. We include nebular emission lines using a grid of CLOUDY simulations \citep{VillaVelez21}.
As output, the code provides a Bayesian-like analysis of each derived parameter.

\subsection{\label{sfh}Star formation history modelling}
To model the SFH, we start from the non-parametric approach\footnote{This {\sc cigale} module is not available in the public version, but can be provided upon request.} described in \cite{Ciesla23} and \cite{ArangoToro23}.
This module makes use of priors to handle the SFR in each time bin of the SFH \citep[see for instance][]{Leja19a,Tacchella20,Suess22}.
In {\sc cigale}, \cite{Ciesla23} implemented and tested this approach using a bursty-continuity prior \citep{Tacchella20}.
With a set of simulated SFH of $z\sim1$ galaxies, they showed that this SFH model provides stellar masses closer to the true ones compared to analytical SFHs as well as more accurate SFRs.
A consistent and extensive analysis using {\sc cigale} fits of galaxies from hydrodynamical simulations will be presented in Arango-Toro et al. (in prep).
However, recently, \cite{Narayanan23} showed that these priors do not allow for the recovering of highly stochastic SFHs expected in galaxies at $z>6$, impacting the measurement of physical properties such as the stellar mass.
Their results showed that a Dirichlet prior, for instance, reaches a limit on the minimum stellar mass that can be derived while other priors, such as a rising one or a prior dedicated to post-starburst galaxies, will provide underestimated stellar masses.
There is a general consensus that different SFH priors will lead to different results and that there is not one prior adapted to all type of galaxies at any redshift \citep[e.g.,][]{Leja19a,Suess22}.
Furthermore, the non parametric SFHs implemented in {\sc cigale} but also in {\sc prospector} \citep{Johnson21} or {\sc bagpipes} \citep{Carnall18} make use of logarithmic binning of time.
Although this binning is adapted for galaxies with ages of a few Gyr, when galaxies are young ($<$1\,Gyr), such an approach might not be needed anymore.

We propose in this work a SFH modelling adapted for early galaxies with ages younger than 1\,Gyr.
We implement, test, and use a non-parametric SFH with linear time binning and a flat prior. 
The aim of this approach is to have an SFH flexible enough to capture possible bursts or stochastic activity, while providing a good estimate of properties like the stellar mass and SFR.
In each bin, except the latest one (that is the closest to the time of observation), the SFR is randomly picked in a uniform distribution ranging from 0 to sfr$_{max}$, an input parameter of the code.
To allow for as much flexibility as possible to recover the right SFR, the latest bin is randomly picked in a log-uniform distribution, ranging from 10$^{-2}$ to sfr$_{max}$.
The assumed SFH is thus a set of 10 bins linearly defined (the age of the galaxy divided in ten equal time bins) except for the latest one which has an imposed duration of 10\,Myr to be able to capture the most recent variation in star formation activity.
This value is an input parameter and can be modified, such as the number of bins.

\begin{figure}[ht] 
 	\includegraphics[width=\columnwidth]{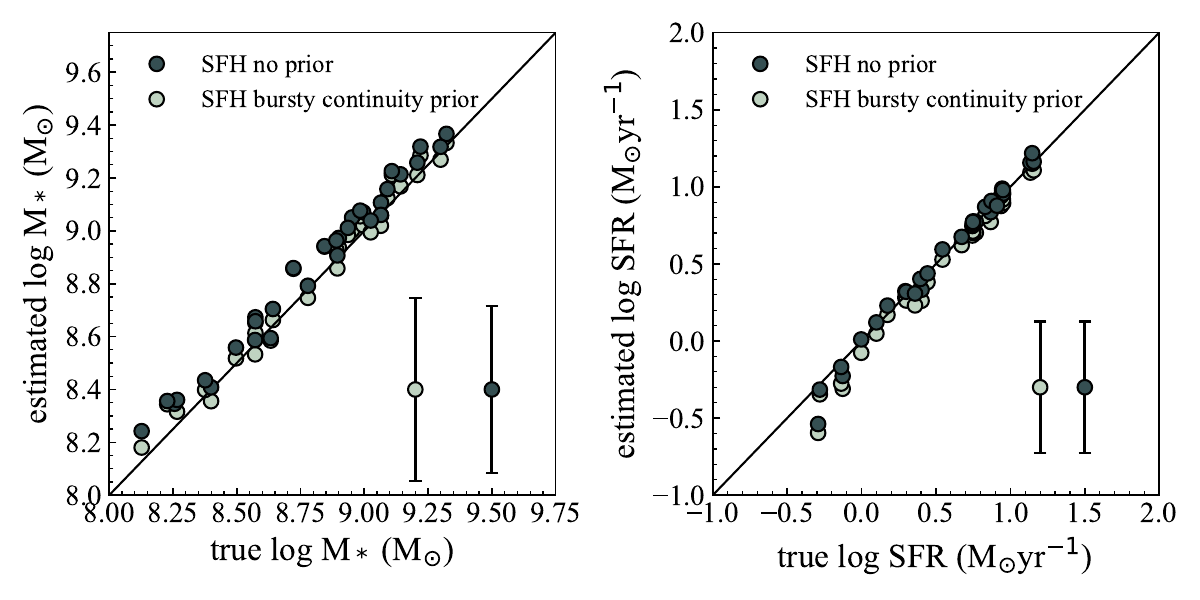}
  	\caption{\label{simu} Left panel: Stellar mass recovered with {\sc cigale} using the non-parametric SFH with no prior versus the true stellar mass of the simulated SEDs. Light green dots are obtained using a non-parametric SFH with bursty continuity prior while the drak green dots are obtained with the non-parametric SFH without prior proposed in this work. Average errors are indicated in the lower right corner of the panel. Right panel: Same as the left panel but for the SFR.}
\end{figure}

\begin{figure*}[ht] 
 	\includegraphics[width=\textwidth]{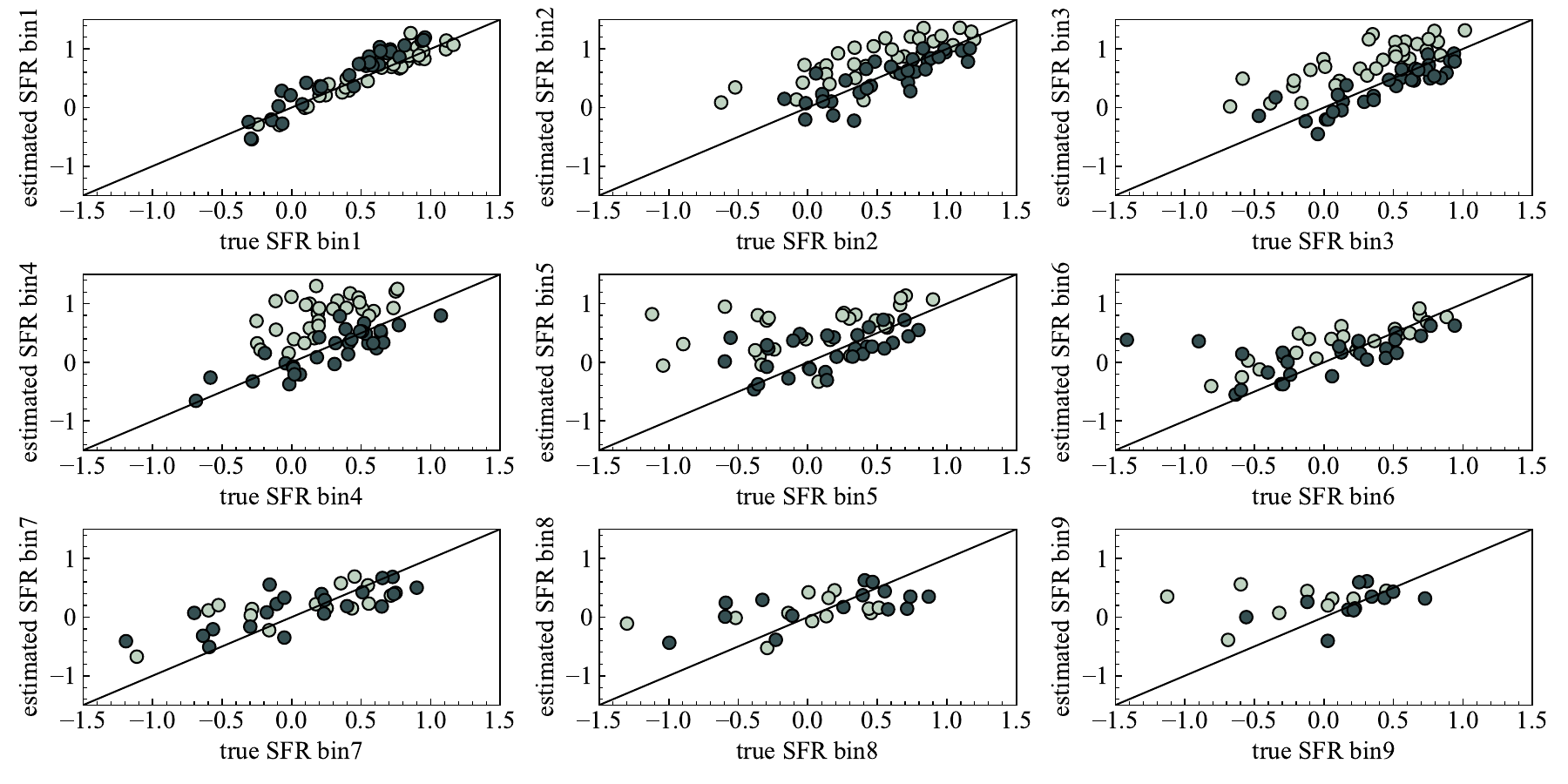}
  	\caption{\label{testsimu} Comparison between the simulated SFHs from \cite{Narayanan23}, binned like {\sc cigale} non-parametric SFHs, and the recovered SFR in the same time bins obtained with {\sc cigale}. Bin 1 corresponds to the latest bin, at time of the observation, while bin 9 corresponds to the earliest time bin. Light green dots are obtained using a non-parametric SFH with bursty continuity prior while the dark green dots are obtained with the non-parametric SFH without prior proposed in this work.}
\end{figure*}

To test our SFH model, we use the set of 32 simulated SFHs presented in \cite{Narayanan23}. They analysed two cosmological simulations at $z\sim7$ with significantly different stellar feedback models, and therefore dramatically different star formation histories.  In detail, they studied the {\sc simba} simulation \citep{Dave19}, as well as a newly run simulation with the  {\sc smuggle} feedback model \citep{marinacci19a}.  The former has a smoother rising star formation history, as is typical in cosmological simulations with artificially pressurised ISM models, while the latter is more bursty owing to the explicit feedback scheme.  \cite{Narayanan23} employed these two models with diverse star formation histories to analyze the ability of non-parametric SFH models in SED fitting in recovering the true stellar masses, and found that outshining from the most recent star formation (at the time of observation) hampered the ability to accurately derive galaxy star formation histories (and therefore, stellar masses) at high-$z$.

We follow the same approach than \cite{Narayanan23} by reducing all sources of uncertainty in the fit to be able to attribute the possible discrepancies only to SFH recovering. 
We thus use {\sc cigale} to create the SEDs from the SIMBA and SMUGGLE simulated SFHs using \cite{BruzualCharlot03} stellar population models and a \cite{Calzetti00} dust attenuation law with an E(B-V)s of 0.08.
These SEDs are integrated into the same set of filters available in JADES to which we associate errors of 20$\%$ for JWST/NIRCam bands and 50$\%$ for HST ones, which are representative of the average errors provided in the JADES catalogue.
We note that {\sc cigale} adds an additionnal 10$\%$ to these values to take into account models uncertainties.

We fit these mocks fluxes with {\sc cigale}. 
This test is purposely simple and designed to identify discrepancies that are only due to SFH recovery.
The output stellar masses and SFRs are compared to the true ones from the simulated galaxies in Fig.~\ref{simu}.
There is a good correlation between recovered stellar masses and the true ones with a small systematic offset of 0.07\,dex.
Instantaneous SFRs are well recovered.
We conducted the same test using the bursty continuity prior and found consistent results, with a good estimates of the stellar masses and SFRs of the SEDs built from the \texttt{SMUGGLE} and \texttt{SIMBA} simulated SFHs.

To evaluate the quality of the SFH reconstruction, we compare the SFR recovered in each time bin to the true ones in Fig.~\ref{testsimu}.
To this aim, we project the SFHs from simulations on the same time grid used by {\sc cigale}.
Results show a better recovery of the SFH by the model with flat prior, compared to the bursty continuity prior, at least up to the fifth bin going in lookbacktime. 
Although both priors provide good estimates of both stellar masses and SFRs, the difference in SFH reconstruction seen in Fig.~\ref{testsimu} is explained by the difference in age estimate provided by these two models.
The bursty continuity prior yields lower formation ages, that is the age at which the galaxy has formed 50\% of its stellar mass.
At earlier time in the life of the galaxies, the SFH is still recovered but the comparison is more dispersed.
We use the results of this test to calibrate our fitting procedure and provide the input set of parameters used to run {\sc cigale} in Table~\ref{inputparam}.
This parameter set results in the best compromise to cover the JADES observed colours while ensuring a reasonable number of models for computing reasons.

\begin{table}
   \centering
   \caption{\cigale\ input parameters used to fit the JADES sample. This set of parameters resulted in 54,000 SED models per redshift.}
   \begin{tabular}{l c l}
   \hline\hline
   \textbf{Parameter} & \textbf{Value} & \\
   \hline
   \multicolumn{3}{c}{\textbf{Non-parametric SFH --} \sc{sfhNlevels$\_$flat}}\\[1mm]  
   $age$ (Gyr) & 3 values depending      & maximum possible\\
               & on redshift bins     & age of galaxies  \\
   sfr$\_$max & 500 M$_{\odot}$yr$^{-1}$ & max. value of the SFR \\[1mm]
   N$_{SFH}$ & 1,000 & \# of SFH per $age$ value \\[1mm]
   Last bin size & 10\,Myr & \\[1mm]
   \multicolumn{3}{c}{\textbf{Emission lines --} \sc{nebular}}\\[1mm]  
   $\log$U      &  -3, -2, -1     & \\
   \multicolumn{3}{c}{\textbf{Dust attenuation --} \sc{dustatt\_modified\_starburst}}\\[1mm]  
   E(B-V)s lines      &  $\big[$0;\ 0.3$\big]$     &6 values linearly sampled  \\
   \hline
   \label{inputparam}
   \end{tabular}
   \footnotetext[1]{Color excess ratio between continuum and nebular emission}
\end{table}

\section{\label{secms}Early galaxies on the SFR-$M_\ast$ plane}

\begin{figure*}[ht] 
 	\includegraphics[width=\textwidth]{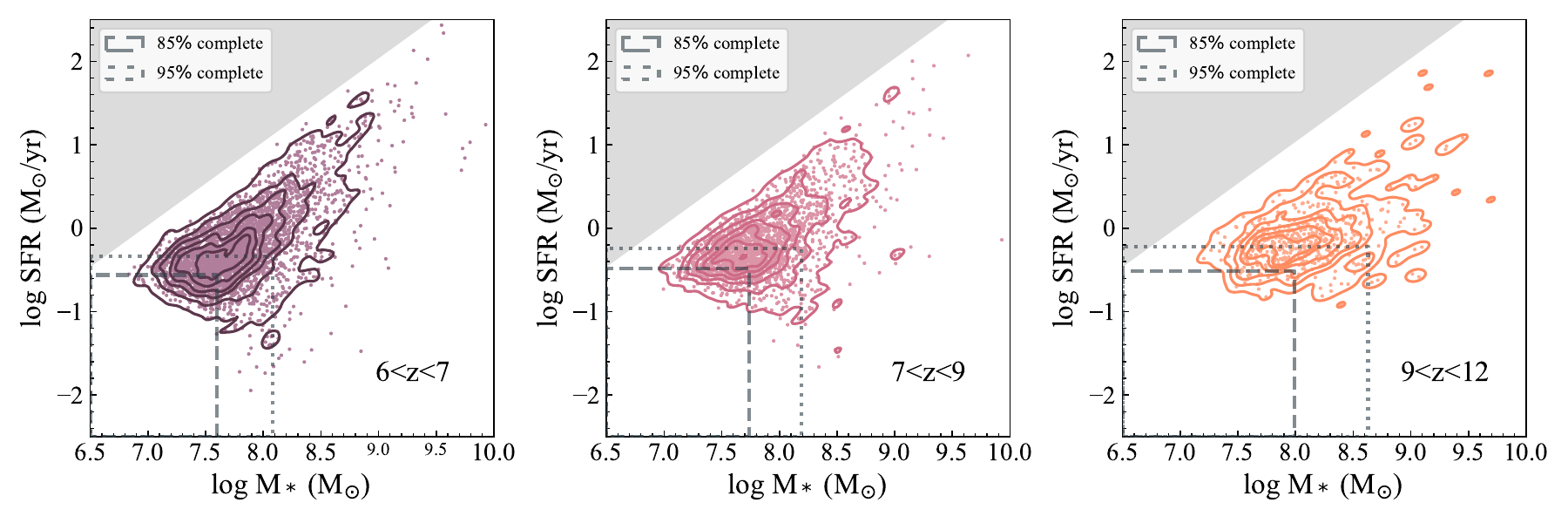}
  	\caption{\label{ms} SFR as a function of $M_\ast$ in three bins of redshift. The positions of JADES galaxies are indicated by the coloured points and contours. The grey dotted and dashed lines indicate where the sample is 95$\%$ and 85$\%$ complete, respectively. The grey regions indicate part of the SFR-$M_\ast$ plane not covered by the models used to do the SED fitting.}
\end{figure*}

\begin{figure}[ht] 
 	\includegraphics[width=\columnwidth]{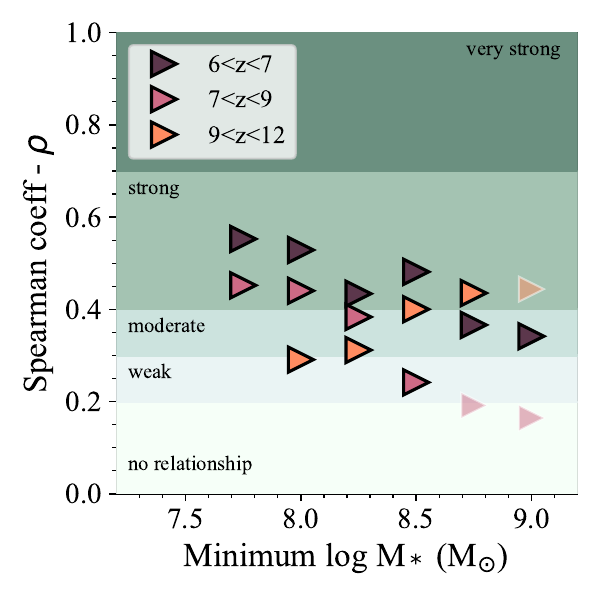}
  	\caption{\label{spearman} Spearman coefficient as a function of the minimum stellar mass used to compute it. Symbol colours indicate the redshift range of the galaxies considered. Those with black contours are considered reliable coefficient value since associated with p-value lower than 0.05. The green shaded regions indicate the interpretation scale.}
\end{figure}

We place the JADES galaxies of our sample in the SFR-$M_\ast$ plane, using the instantaneous SFR and separating the three redshift bins (Fig.~\ref{ms}).

\subsection{Sample completeness}
To determine the mass completeness of the sample studied in this work, we use the method described in \cite{Pozzetti10} \citep[see also][]{Florez20,Mountrichas22}.
We calculate the limiting stellar mass, that is the mass that a galaxy would have if its magnitude would be equal to the limiting magnitude of the survey: $\log$M$_{\ast,lim}$$=\log$M$_* + 0.4\times (m_{AB} - m_{AB,lim})$.
For this, we use the NIRCam/F200W filter and the flux sensitivity of 4.4\,nJy provided in \cite{Rieke23}.
We then select the 20$\%$ faintest galaxies of the sample and calculate the limiting masses at the 95$^{th}$ and 85$^{th}$ percentiles of this distribution.
The sample is 95$\%$ complete above $\log$(M$_{*,lim}$/M$_{\odot}$)= 8.08, 8.19, and 8.63, at $6<z<7$, $7<z<9$, and $9<z<12$, respectively, and 85$\%$ complete above $\log$(M$_{\ast,lim}$/M$_{\odot}$)= 7.60, 7.74, and 7.99, in the same redshift bins.
As a second step, we compute the SFR above which our sample is complete, using the same procedure than for the stellar mass.
The 95$\%$ completeness is reached at $\log$(SFR$_{lim}$/M$_{\odot}$yr$^{-1}$) are -0.34, -0.24, -0.22, at $6<z<7$, $7<z<9$, and $9<z<12$, respectively, while the 85$\%$ completeness is reached at -0.56, -0.48, and -0.51.
These limits are indicated on Fig.~\ref{ms}.

\subsection{SFR-$M_\ast$ correlation}
In the $6<z<7$ redshift bin, there is evidence for a relation between the SFR and stellar mass which is strengthened by the absence of galaxies with strong SFR compared to the main population.
This limit is not induced by the models used to do the SED fitting that allow to reach all the white part of the SFR-$M_\ast$ planes shown in Fig.~\ref{ms}.
In the $7<z<9$ redshift bin, we see a relation between the SFR and the stellar mass that is broader than in the previous redshift bin. 
Finally, in the highest redshift bin, $9<z<12$, there is no clear evidence for a relationship between SFR and $M_\ast$.
Instead, although showing some dispersion, the SFR seems to be constant as a function of stellar mass. 
Only at high masses ($\log$$M_\ast$$\gtrsim$9), the SFR of the few galaxies at these masses seems to increase with $M_\ast$.
At these high redshifts, the SFR seems to be insensitive to the stellar mass of the galaxy.
To test if this effect could be due to a bias in estimating the stellar mass, given the rest-frame wavelength coverage of our sample, we remove the longest NIRCam filters of the SEDs of $z$$\sim$8 galaxies to mimic the rest-frame coverage of the galaxies in the highest redshift bin.
The stellar masses obtained with this narrower wavelength range are consistent with those obtain with the full filter set, with a ratio of 1.10$\pm$0.23 between the two stellar masses estimates.
We conclude that, with the data in hand, the constant SFR distribution observed as a function of stellar mass at $9<z<12$ is not due to stellar mass bias linked to rest-frame wavelength coverage. 

To characterise the relationship between the SFR and stellar mass seen in Fig.~\ref{ms}, we compute the Spearman coefficient as an indicator of its strength.
We vary the minimum stellar mass considered in the subsamples and show the results of this test in Fig.~\ref{spearman}, considering only Spearman coefficients estimated with a p-value lower than 0.05.
Using the different mass thresholds, the derived coefficients range from 0.15 to 0.5.
At $6<z<7$, the Spearman coefficient, when considering the sample above $\log$$M_\ast$$=$7.5, is 0.48 corresponding to a strong relationship between the SFR and stellar mass parameters \citep{DanceyReidy04}.
At $7<z<9$, the results indicate a moderate relationship when considering the same threshold. 
For the highest redshift bin ($9<z<12$), the Spearman coefficients values only indicate moderate or strong relationship between the two parameters for high masses, that is above $\log$$M_\ast$$=$8.5.
At these high redshifts, when considering all sources, there is only weak evidence for it.
The observations made on Fig.~\ref{ms} are confirmed by the Spearman coefficient analysis, highlighting a strong relationship between the SFR and stellar mass at $z<7$ that slightly weakens at $7<z<9$.
At higher redshifts, the relation is strong only for the most massive galaxies.

\begin{figure}[ht] 
 	\includegraphics[width=\columnwidth]{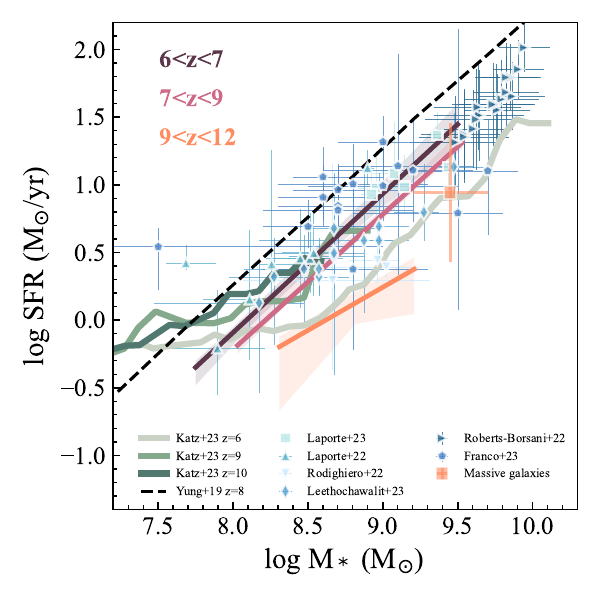}
  	\caption{\label{ms_lit} SFR as a function of stellar mass. The relations found in the redshift bins studied in this work are shown with the solid lines and shaded regions. They are compared to samples of the literature. Relations from simulations are shown in dashed black line \citep{Yung19} and green solid lines \citep{Katz23}. SFR and masses from the literature have been converted to \cite{Salpeter55} IMF dividing by 0.63 from a \cite{Chabrier03} IMF or 0.67 from a \cite{KroupaBoily02} one \citep{MadauDickinson14,Bernardi18}.}
\end{figure}

We compare our SFR-$M_\ast$ relations with samples of early galaxies available in the literature in Fig~\ref{ms_lit}, converting all masses and SFR to \cite{Salpeter55} IMF.
There is a good agreement between our relations and observed sources presented in the literature when comparing with our $6<z<7$ and $7<z<9$ redshift bins, although we estimate slightly lower SFR at a given mass.
This can be due to different selections between our sample and those of the literature that were generally obtained through specific criteria (colour diagrams, spectroscopic observations, etc).
However, we observe a flatter relation for our highest redshift bin ($9<z<12$) which falls off with a slope of 0.64.
Although built with little statistic, we indicate the median SFR of galaxies with $\log$$M_\ast$$\sim$9.5 as an indication but we do not use this point in the fit.
We add for comparison predictions from the models of \cite{Yung23} and the \texttt{Sphinx} simulations \cite{Katz23}.
In the case of the \texttt{Sphinx} simulations, we note that, above $\log$$M_\ast$$=$8.5, the normalisation of their relation increases with increasing redshift while we observe the opposite with our data. 

\section{\label{sectrans}From stochastic to continuous SFH}

In this section, we aim at characterising the SFH of galaxies and refer as stochastic a SFH where the SFR at a given time bin, defined in Sect.~\ref{sedfit}, is independent from the SFR of the previous time bin or in other words, can not be predicted from the SFR of the previous bin. 
With this stochasticity, we expect to see galaxies going up and down on the SFR-$M_\ast$ plane with no coherent movement.
This is opposed to the concept of secularity implied by the MS for which knowing the SFR at a given stellar mass allows you to predict the SFR of the next stellar mass bin.
When on the MS, we still expect galaxies to exhibit a certain level of stochasticity \citep[e.g.][]{Tacchella16,Ciesla23} but within the common flow that builds the MS.

\begin{figure}[ht] 
 	\includegraphics[width=\columnwidth]{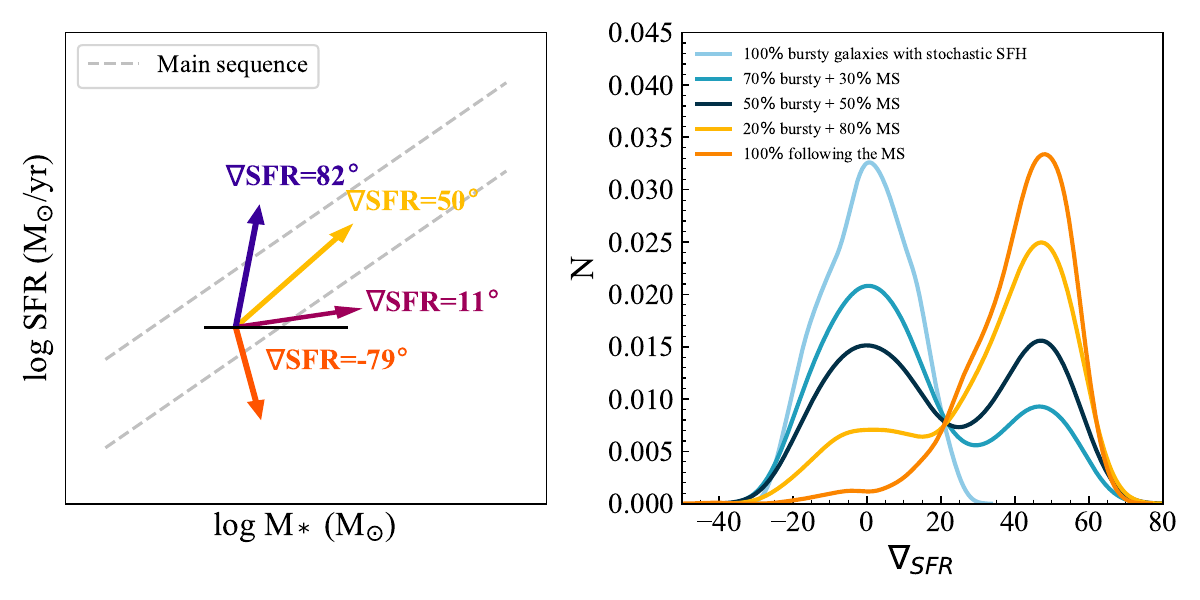}
  	\caption{\label{gradexamp} Left panel: Schematic example of the definition of $\nabla$SFR. The gradient corresponding to each arrow is indicated with the same color and show the path that galaxies followed recently. The point of the arrow indicate the position of the galaxy when observed. The black line is the reference from which the angle of the gradient is computed. The grey dashed lines mark the position of the MS with its the dispersion. Right panel: Mock distributions of $\nabla$SFR obtained from simple modelling. }
\end{figure}

\begin{figure*}[ht] 
 	\includegraphics[width=0.33\textwidth]{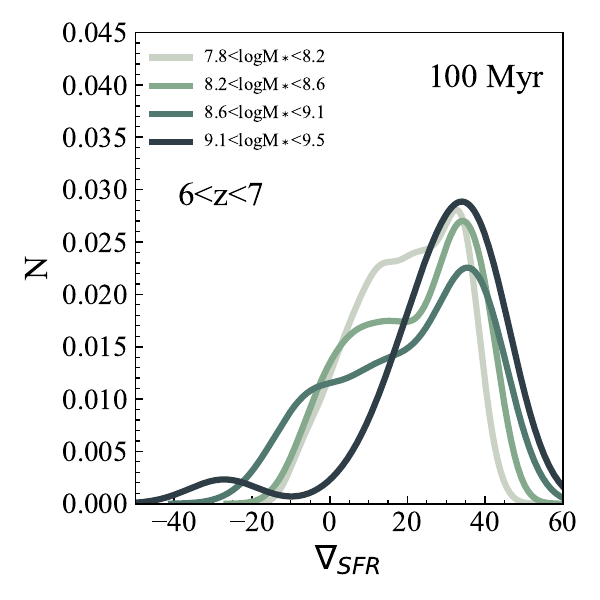}
 	\includegraphics[width=0.33\textwidth]{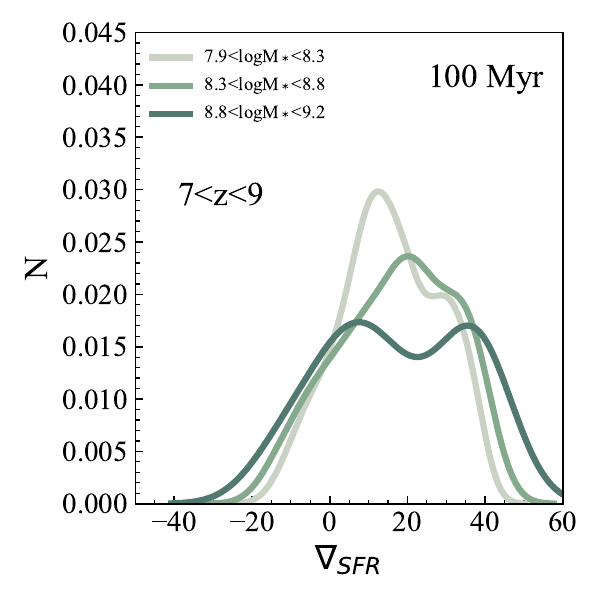}
 	\includegraphics[width=0.33\textwidth]{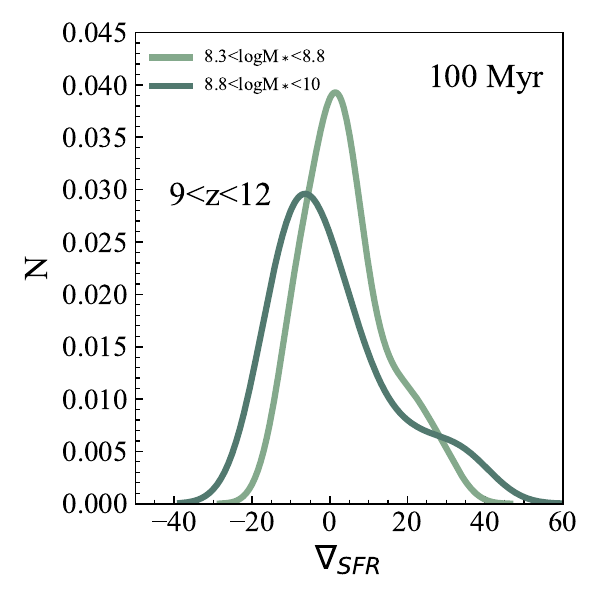}
  	\caption{\label{trans} Distribution of $\nabla$SFR computed over 100\,Myr in the three redshift bins. The different colors indicate different stellar mass bins.}
\end{figure*}

We use the SFR gradient ($\nabla$SFR) proposed in \cite{Ciesla23} to probe the movement of a galaxy in the SFR-$M_\ast$ plane and provide a dynamical approach of the MS.
The SFR gradient is an angle (degree) relative to a constant SFR line in the SFR-$M_\ast$ plane, in log space, that can be computed over different timescales. 
Tested and used in \cite{Ciesla23} and \cite{ArangoToro23}, it is an indicator of the recent SFH of galaxies.
As shown in the left panel of Fig.~\ref{gradexamp}, a galaxy with a strong positive $\nabla$SFR is undergoing a starburst event and is going up on the SFR-$M_\ast$ plane while a galaxy with a strong negative gradient is undergoing a rapid and drastic quenching and is going down in the plane.
This parameter, like the other parameters derived by {\sc cigale}, is estimated through PDF analysis.

To help with the interpretation of $\nabla$SFR and its distribution from a galaxy population, we compute mock $\nabla$SFR assuming two cases.
The first one is when galaxies have a completely stochastic SFH. 
We thus assume a flat prior on the SFR that a galaxy with $\log$M$_*$$=$8 can reach between $-0.2<\log$SFR$<0.5$, compatible with the SFR spread observed in Fig.~\ref{ms} at this stellar mass.
We show the resulting distributions in the right panel of Fig.~\ref{gradexamp}.
The distribution of galaxies with a completely stochastic SFH, within the boundaries set in our assumptions, is centered around 0\degre and spreads between -20\degre and 20\degre.
In the second case, we assume that the SFR of the mock galaxies can only follow the relation characterised in Fig.~\ref{ms_lit} assuming a 0.5\,dex dispersion.
In this case, a slightly narrower distribution is observed peaking around $\nabla$SFR$=$45\degre.
The distributions of these two case scenarios have thus very different characteristics.
In the same figure, we show composite distributions to illustrate how mixed populations (composed of both stochastic and secular SFH galaxies) affects the gradient distributions.

We now compute the gradient distribution of our JADES galaxies to probe the variety of their SFHs.
In Fig.~\ref{trans}, we show this distribution computed over the last 100\,Myr in each redshift bin, and for different stellar mass bins.

At $6<z<7$, almost all galaxies show a peak around 40$\degre$ that is compatible with what we expect from galaxies following the flow of MS.
However, galaxies with $\log$$M_\ast$ between 7.8 and 8.2 have a broader distribution of $\nabla$SFR from -20\degre to 45\degre.
This indicates that a large fraction of low mass galaxies tends have stochastic SF activity.
With increasing stellar mass, the broad distribution changes with the increasing dominance of the peak at high positive values of $\nabla$SFR.
For the highest stellar mass bin ($\log$$M_\ast$ between 9.1 and 9.5), the distribution has one peak around 40\degre and is slightly skewed towards lower gradient values.
According to the mock distributions of Fig.~\ref{gradexamp}, this indicates a dominance of galaxies with secular SF that follow the flow of the MS, discussed in the previous section, with a small contribution from galaxies with stochastic SFH.

We repeat this analysis in the two other reshift bins.
The $7<z<9$ bin reveals that the galaxy population shows a broad, double peaked, distribution of $\nabla$SFR.
For the highest mass bin ($\log$$M_\ast$ between 8.6 and 9.1) the peak at high $\nabla$SFR values, around 40\degre, is seen but less prominent than in the lower redshift bin.
A second peak, around 5\degre, appears which is consistent with a stochastic activity as shown in Fig.~\ref{gradexamp}.
In the lowest mass bin, the same type of distribution with the peak around low gradient values ($\sim$10\degre) starts to be dominant compared to the MS gradient peak around 40\degre.
As observed in the lower redshift bin, at low mass, the SFH of galaxies is predominantly stochastic.

Finally, in the $9<z<12$ redshift bin, the $\nabla$SFR distribution of the two stellar mass bins is clearly peaked around 0$\degre$ with a weak secondary peak around 30-40\degre.
The dominant peak is clearly compatible with a stochastic SFH drawn from a narrow SFR range as shown in purple in Fig.~\ref{gradexamp}.
This strongly highlights that the dominant SFH type is stochastic at this redshift with a small part of massive galaxies that are compatible with secular SF mode.

In Appendix~\ref{gradsnr}, we show how the results deduced from Fig.~\ref{trans} would be affected by degrading the constraint on $\nabla$SFR. 
The main characteristics discussed in this section are recovered even assuming an error following a normal distribution with $\sigma_{\nabla SFR}=20\degre$ (Fig.~\ref{trans_err}).

In the gradient distribution of Fig.~\ref{trans}, the transition between the stochastic and continuous SFH seems to happen at stellar masses lower than 8.6 in the $6<z<7$ and $7<z<9$ redshift bins.
This transition mass is expected from simulations.
For instance, in their sample of galaxies from FIRE-2 zoom-in simulations \citep{Hopkins18}, \cite{Ma18} observed a large scatter in their MS relation at masses lower than $\log$$M_\ast$=8 (with \cite{Kroupa01} IMF, thus $\sim$8.2 in \cite{Salpeter55} IMF) due to stronger burstiness in the SFHs.
Consistent results obtained from simulations were also found by \cite{Dayal13} and \cite{KimmCen14}.
Using observations of local analogues of early galaxies, \cite{Emami19} studied the distribution of H$\alpha$ to UV luminosity ratio as a proxy of burstiness and found its distribution to be larger in galaxies with $\log$($M_\ast$/M$_{\odot})$$<$8 (with \cite{Chabrier03} IMF, thus $\sim$8.2 in \cite{Salpeter55} IMF).
These transition masses are consistent with what we observe with the JADES sample for which the median error of the stellar mass is 0.4\,dex.
Above $\log$$M_\ast$$\sim$9, galaxies with redshift between 6 and 7 show SFHs that are compatible with the majority of them following the MS.
\section{\label{discussion}Discussion}

\subsection{Star formation mode}
\begin{figure}[ht] 
 	\includegraphics[width=\columnwidth]{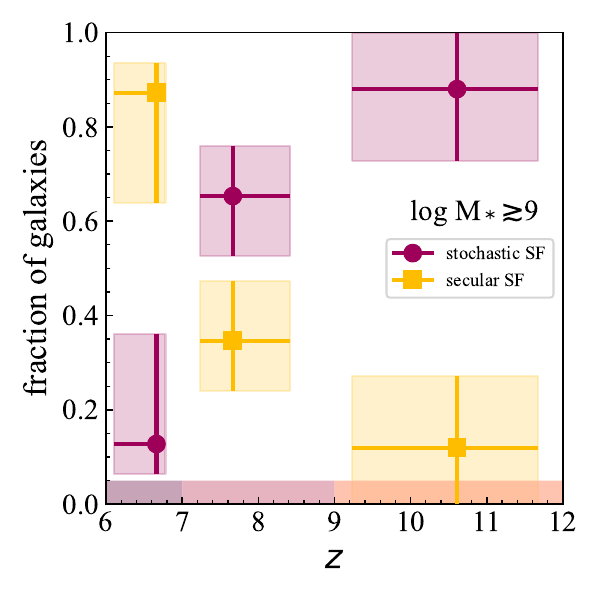}
  	\caption{\label{gradfrac} Fraction of galaxies in the secular SFH gradient distribution (peak around 40\degre, yellow squares) and in the stochastic SFH distribution (peak around 0\degre, purple circles) as a function of redshift.}
\end{figure}

Several studies, based on both simulations \citep[e.g.,][]{Dayal13,KimmCen14,Ma18,Legrand22,Gelli23,Sun23,Wilkins23} and observations \citep[e.g.,][]{Faisst19,Emami19,Dressler23b,Looser23b}, have shown that early galaxies had bursty or stochastic SFHs.
In this work, we reach a consistent conclusion for galaxies at $z\gtrsim9$ through the analysis of their SFR gradient computed over the last 100\,Myr.
At lower redshift, we see the emergence of galaxies with a gradient distribution compatible with a rising SFH and with a coherent mouvement of galaxies following the flow of a main sequence.
To understand this transition between the star-formation mode of galaxies at $z\sim9$, we show in Fig.~\ref{gradfrac} the fraction of massive ($\log$M$_*$$>$9) galaxies that have a gradient compatible with a stochastic SFH and the fraction of massive galaxies following the flow of the MS with a gradient distribution peaking aroung 30-40\degre.
There is a clear transition occurring between $z$$=$7 and $z$$=$9, thus in $\sim$200\,Myr, where the majority of galaxies have a stochastic SFH at $z>9$ (more than 80\%) while at $z$$\lesssim$7 galaxies having a secular star-formation mode dominates the population of massive galaxies.
In Fig.~\ref{gradfrac_err}, we show how the results on Fig.~\ref{gradfrac} are affected by a weaker constraint on the $\nabla$SFR estimates and found no differences in our conclusions.

\subsection{Impact on UV emission and variability}

JWST Early Release Science results have highlighted an overabundance of galaxies at $z$$=$10 and higher, probed by the UVLF \citep[e.g.,][]{Harikane23,Finkelstein23}.
Several explanations have been proposed to understand this overabundance with physical processes involving the median level of UV emission \citep{Inayoshi22,Ferrara22,Dekel23,Mason23,Yung23} or its stochasticity \citep{Mason23,MirochaFurlanetto23,Shen23}.

The stochasticity of star formation has been quantified in the literature with $\sigma_{UV}$, the dispersion in mag around a median M$_{UV}$ obtained from a M$_{UV}-$M$_{halo}$ relation.
\cite{Shen23} proposed three sources of variability that would impact $\sigma_{UV}$ and that could, all together, contributes significantly to this excess of UV emission seen in the UVLF.
They first consider the halo assembly, that is the mass accretion rate of dark matter halos that follows a log-normal distribution with a typical $\sigma_{DM}$ of $\sim$0.3\,dex \citep[e.g.,][]{MirochaFurlanetto23}.
Then, they consider the effect of dust attenuation, either through geometrical distribution or the effect of radiative feedback from star formation burst on the dust content \citep[e.g.,][]{Ferrara22}. 
Based on their adopted A$_{UV}$-$\beta$ relation they estimate dispersion due to dust, $\sigma_{dust}$, of $\sim$0.7\,dex.
Finally, the third possible cause of variability is star formation activity such as the one discussed in our study and they adopt a conservative value of $\sigma_{SF}$ of $\geq$0.3\,dex corresponding to the scatter of the MS at lower redshift.
The three $\sigma$ considered together constitute $\sigma_{UV}$ that they estimate to vary from 0.75\,mag at the bright end of the UVLF to 1.5\,mag at the faint end, at $z$$\sim$9.

\begin{figure}[ht] 
 	\includegraphics[width=\columnwidth]{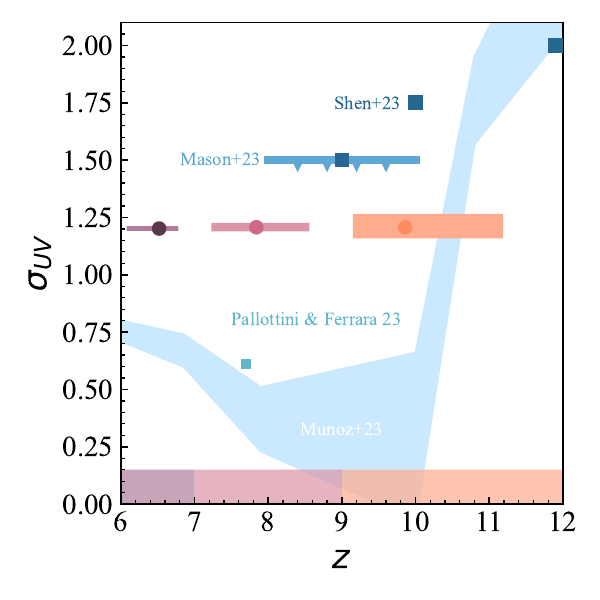}
  	\caption{\label{sigmauv} $\sigma_{UV}$ as a function of redshift. Estimates from the literature, based on simulation/theoretical approaches are shown in blue colours.}
\end{figure}

Using our sample, we derive the dispersion of the absolute magnitude M$_{UV}$ distribution, $\sigma_{UV}$, computed from a rest-frame wavelength window between 145 and 155\,nm using our SED fitting run.
Our derived $\sigma_{UV}$ are shown in Fig.~\ref{sigmauv} as a function of redshift.
Considering that stellar mass traces the halo mass, we compute the M$_{UV}$ distribution in between $\log$($M_\ast$/M$_{\odot})$$=$8 (8.5 at $9<z<12$), the 95\% completeness mass limit, and $\log$M$_*=$9.5.
These results in UV magnitude ranges are -16.5$>$M$_{UV}$$>$-18.9, -16.3$>$M$_{UV}$$>$-18.6, and -16.6$>$M$_{UV}$$>$-18.4, at $6<z<7$, $7<z<9$, and $9<z<12$, respectively. 
In all redshift bins, we find $\sigma_{UV}$ of 1.2$\pm$0.1\,mag.
Subdividing the stellar mass bins results in consistent estimates.

Following the same approach than \cite{Shen23}, we decompose our $\sigma_{UV}$ in the three components. 
For the variation of DM accretion rate, we cannot probe it with our data and thus use the same value than \cite{Shen23}, that is 0.3\,dex.
For the effect of dust, we derive from our SED modelling the distribution of attenuation in the same rest-frame band used to derive M$_{UV}$ and obtain $\sim$0.35\,mag, depending on the redshift bin.
Finally, we have estimated the observed dispersion of the galaxies in the SFR-$M_\ast$ plane and obtained $\sim$0.5\,dex in the considered stellar mass bin.
Combining these, we obtain a $\sigma_{UV}$ of 0.68\,mag which is lower than our observed M$_{UV}$ based $\sigma_{UV}$. 

Our $\sigma_{UV}$ are compared to estimates from the literature in Fig.~\ref{sigmauv}.
\cite{PallottiniFerrara23} used the {\sc serra} simulations and estimated a $\sigma_{UV}$ of 0.61\,mag from a sample of 245 simulated galaxies at $z$$\sim$7.7 studying the time-dependant variation of the SFR over average SFR, an indicator of burstiness.
They concluded that, from their simulations, the SFR variations cannot account for the required $z$$\gtrsim$10 boost in UVLF.
Although their estimated $\sigma_{UV}$ is lower than ours, our analysis reaches concistent conclusions.
\cite{Munoz23} used semi-empirical models that allowed to derive the expected evolution of $\sigma_{UV}$ finding low values ($<$0.8\,mag) up to $z$$\sim$10, followed by a strong and rapid increase up to 2\,mag at $z$$\sim$11.
There is a strong increase of their predicted $\sigma_{UV}$ over the $9<z<12$ range than could be compatible with our estimate.
With a different approach, \cite{Mason23} used empirical models to study the distribution of the M$_{UV}-$M$_{halo}$ relation and showed that young-aged galaxies significantly upscatter UV emission up to 1.5\,mag above the median relation.
Although within the range proposed by \cite{Mason23}, our $\sigma_{UV}$ values derived from observations and SED modelling are lower than the estimates from \cite{Shen23}.
This level of stochasticity is not sufficient to explain the excess of UV emission seen from the UVLF.

\section{\label{conclusions}Conclusions}

We have used a subsample of galaxies at $6<z<12$ from the JADES public catalogue \citep{Rieke23} to investigate the variability of star formation at these redshifts.
As the star-forming activity is expected to be highly stochastic at these high redshifts, we have implemented in {\sc cigale} a non-parametric SFH built with a linear sampling of time and no imposed prior.
Tested on the simulated galaxies studied in \cite{Narayanan23}, we have found that although the usual bursty continuity prior recovers well the stellar mass and SFR of the simulated galaxies, the proposed parametrisation using linear bins and no prior provide a better reconstruction of the SFH.
We therefore used this approach, fitted the JADES galaxies with {\sc cigale}, and reached the following conclusions:

\begin{itemize}
    \item We observe a strong relationship between the SFR and stellar mass in the 6$<z<$7 redshift range that weakens in the 7$<z<$9 redshift bin. There is no strong evidence for a link between SFR and stellar mass at $z$$>$9 for low mass galaxies with, instead, a rather flat SFR distribution as a function of stellar mass. The few massive ($\log$$M_\ast\gtrsim$9) galaxies at this high redshift bin gives a hint for a stronger relationship between SFR and $M_\ast$.
    
    \item We use the SFR gradient, an indicator of the movement of galaxies in the SFR-$M_\ast$ plane and thus directly linked to galaxies' SFH. Computed over 100\,Myr, the gradient distribution show an evolution with redshift for the most massive galaxies ($\log$$M_\ast\gtrsim$9): the peaked distribution centered around 40\degre, expected from a secular star-formation activity of galaxies following the flow of the MS, fades with increasing redshift whereas the peaked distribution centered around 0\degre, consistent with a stochastic SFH, increases with increasing redshift. The transition between the prominence of secular star-formation mode and the stochastic SFH mode occurs in the 7$<z<$9 redshift bin.
    
    \item At $6<z<9$, low mass galaxies ($\log$M$_*$$<$8.5) have stochastic SFH.

    \item From our SED modelling, we derive the dispersion of the UV absolute magnitude M$_{UV}$ distribution and obtain $\sigma_{UV}$ of 1.2$\pm$0.1\,mag in the three redshift bins. This value seems to be too low to explain the UV excess of the UVLF at high redshift from stochasticity of star-formation only.
    
\end{itemize}

We observe a change in the star formation history of galaxies that is marked around $z$$\sim$9, i.e., the redshift range where the JWST makes a noticeable difference compared to previous instruments. In the most recent period ($z$$<$9), galaxies have been found to dominantly form their stars in a secular star formation mode, i.e. with moderate oscillations of their SFR and with a strong correlation with their stellar mass. This mode breaks down around 550\,Myr after the big bang ($z$$\sim$9). At earlier epochs, a more stochastic mode of star formation dominates with on and off star-formation phases until galaxies reach a critical mass and enter in a more secular star-formation phase putting them on the star-formation main sequence. At $z<7$, about 85\% of the galaxies have reached the main sequence.

We note that despite this early stochastic period, the level of variation that we estimate remains below the very high stochasticity that would be needed to explain the excess of galaxies found in the UV luminosity function at $z$$\gtrsim$10.

We recall, however, that our results are based on a limited rest-frame wavelength coverage at these very high redshifts where they rely on the quality of photometric redshifts. Future observations with the JWST will provide more material to investigate and quantify this transition further.

\begin{acknowledgements}
L.~C. thanks M.~Boquien, as always, for useful discussions on {\sc cigale}.
The authors warmly thank the JADES team for the huge work and effort put in the preparation, observation, and production of the data used in this work.
This project has received financial support from CNRS and CNES through the MITI interdisciplinary programs.
We acknowledge the funding of the French Agence Nationale de la Recherche for the project iMAGE (grant ANR-22-CE31-0007).
C.~G.~G. acknowledges support from CNES.
NIRCam was built by a team at the University of Arizona (UofA) and Lockheed Martin's Advanced Technology Center, led by Prof. Marcia Rieke at UoA.
This work is based on observations made with the NASA/ESA/CSA James Webb Space Telescope. 
The data were obtained from the Mikulski Archive for Space Telescopes at the Space Telescope Science Institute, which is operated by the Association of Universities for Research in Astronomy, Inc., under NASA contract NAS 5-03127 for JWST. 
\end{acknowledgements}
\bibliographystyle{aa}
\bibliography{JADES_SFH_ZGT6/jades_SFH_zgt6}
\appendix
\section{\label{gradsnr}Impact of the constraint on $\nabla$SFR on the results}
\begin{figure*}[ht] 
 	\includegraphics[width=0.33\textwidth]{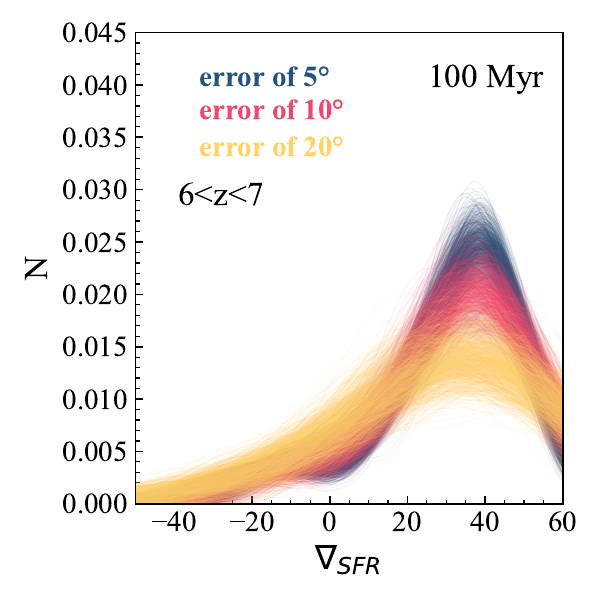}
 	\includegraphics[width=0.33\textwidth]{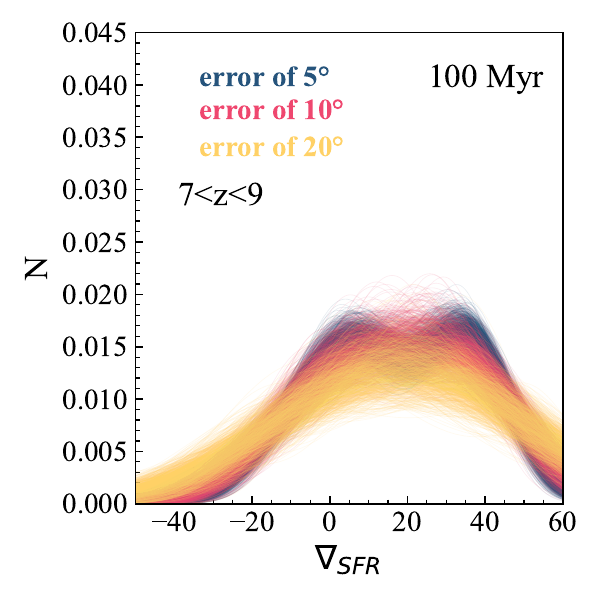}
 	\includegraphics[width=0.33\textwidth]{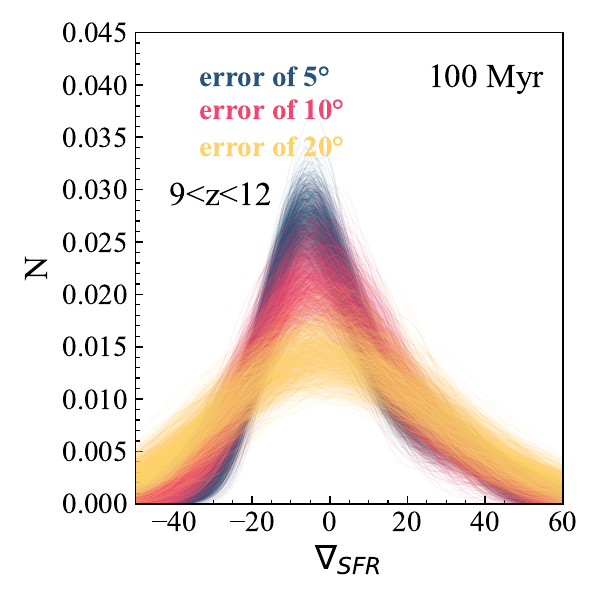}
  	\caption{\label{trans_err} Distribution of $\nabla$SFR computed over 100\,Myr in the three redshift bins. The different colors indicate different SNR assumed to perturb the gradient distributions.}
\end{figure*}

\begin{figure}[ht] 
 	\includegraphics[width=\columnwidth]{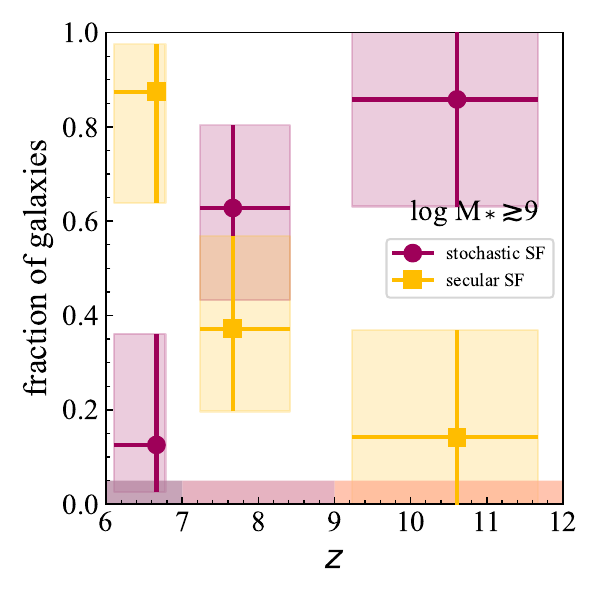}
  	\caption{\label{gradfrac_err} Fraction of galaxies in the secular SFH gradient distribution (peak around 40\degre, yellow squares) and in the stochastic SFH distribution (peak around 0\degre, purple circles), assuming that these distribution are normal, as a function of redshift.}
\end{figure}

To understand how a weaker constraint of the SFR gradient would impact our results of Fig.~\ref{trans}, we perturb the gradient estimates of the highest mass bin of each redshift range a thousand times, assuming different errors on $\nabla$SFR.
We perturb $\nabla$SFR estimates by randmly peaking a value in a normal distribution with a $\sigma$ of 5\degre, 10\degre, or 20\degre.
As shown in Fig.~\ref{trans_err}, the distributions broaden with increasing $\sigma$ but the position of the peaks is still clear and allow to observe the two type of distributions that we interpret as galaxies with a stochastic SFH ($\nabla$SFR$\sim$0\degre) and galaxies with a SFH following a coherent movement, indicative of secular star-formation mode ($\nabla$SFR$\sim$40\degre), even with $\sigma=$20\degre.
We repeat the same exercise to test the results of Fig.~\ref{gradfrac} and reach the same conclusion (Fig.~\ref{gradfrac_err}).
Even considering a high level of error of the estimate of $\nabla$SFR, the transition between stochastic and secular star-formation modes is still clear.

\end{document}